\documentclass[preprint]{iucr}
\journalcode{A}

\usepackage{graphicx}
\usepackage{xspace}
\usepackage{threeparttable}
\usepackage{geometry}
\usepackage{color,xcolor}
\usepackage{epstopdf}
\usepackage{mathtools, amsmath, amsthm, amssymb, amsfonts,bm}
\usepackage[12pt]{moresize}
\usepackage{multirow}
\usepackage{url}
\usepackage{ulem}

\usepackage{threeparttable}
\usepackage{xparse}
\usepackage{verbatim}

\graphicspath{{./figures/}}

\definecolor{ykcolor}{HTML}{4d1063}

\definecolor{citationtags}{HTML}{21a889}

\definecolor{dgreen}{HTML}{008000}

\renewcommand{\vec}[1]{\ensuremath\mathbf{#1}}
\newcommand{\be}{\begin{enumerate}[wide, labelwidth=!, labelindent=0pt,
        label=\textbf{\textcolor{blue}{\arabic*}.}]}
    \newcommand{\bei}{\begin{enumerate}}
        \newcommand{\ee}{\end{enumerate}}
    \newcounter{saveenumi}

\newcommand{\insitu}{{\it in~situ}}

\newcommand{\nba}[1]{}

\newcommand{\qmax}{\ensuremath{Q_{\mathrm{max}}}\xspace}
\newcommand{\qmin}{\ensuremath{Q_{\mathrm{min}}}\xspace}
\newcommand{\rmin}{\ensuremath{r_{\mathrm{min}}}\xspace}
\newcommand{\rmax}{\ensuremath{r_{\mathrm{max}}}\xspace}

\newcommand{\cmi}{\textsc{Diffpy-CMI}\xspace}

\newcommand{\xpdacq}{\textsc{xpdAcq}\xspace}
\newcommand{\xpdan}{\textsc{xpdAn}\xspace}
\newcommand{\rapidz}{\textsc{rapidz}\xspace}
\newcommand{\sklearn}{\textsc{scikit-learn}\xspace}
\newcommand{\norm}[1]{\left\lVert#1\right\rVert}
\newcommand{\demourl}{https://github.com/GENESIS-EFRC/streaming-matrix-factorization}
\newcommand{\pdfitcurl}{https://pdfitc.org/}

\makeatletter
\newcommand{\floatcaption}{%
    \ifx \@captype \@undefined \@latex@error {\noexpand \caption outside float}\@ehd \expandafter \@gobble \else \refstepcounter \@captype \expandafter \@firstofone \fi {\@dblarg {\@caption \@captype }}%
}%
\makeatother

\begin{document}

\title{Validation of non-negative matrix factorization for assessment of atomic pair-distribution function (PDF) data in a real-time streaming context}
\author[a]{Chia-Hao}{Liu}
\author[a]{Christopher J.} {Wright}
\author[a]{Ran}{Gu}
\author[a]{Sasaank}{Bandi}
\author[b]{Allison}{Wustrow}
\author[b]{Paul K.}{Todd}
\author[c]{Daniel}{O'Nolan}
\author[c]{Michelle L.}{Beauvais}
\author[b]{James R.}{Neilson}
\author[c]{Peter J.}{Chupas}
\author[c]{Karena W.}{Chapman}
\author[a,d]{Simon J. L.}{Billinge}
%\email{sb2896@columbia.edu}
\aff[a]{Department of Applied Physics and Applied Mathematics, Columbia University, \city{New York}, NY 10016, \country{USA}}
\aff[b]{Department of Chemistry, Colorado State University, \city{Fort Collins}, Colorado 80523, \country{USA}}
\aff[c]{Department of Chemistry, Stony Brook University, \city{Stony Brook}, New York 11794, \country{USA}}
\aff[d]{Condensed Matter Physics and Materials Science Department, Brookhaven National Laboratory,\city{Upton}, New York~11973, \country{USA}}
\date{\today}
%\author[a]{Sasaank}{Bandi}
%\author[a]{Chia-Hao}{Liu}
%\author[a]{Christopher J.}{Wright}
%\author[a]{Ran}{Gu}
%\author[a]{Simon J.~L.}{Billinge}
%\aff[a]{Department of Applied Physics and Applied Mathematics, Columbia University, \city{New York}, NY 10016, \country{USA}}
%\aff[b]{Condensed Matter Physics and Materials Science Department, Brookhaven National Laboratory, Upton, NY 11973}

\date{\today}
\maketitle
\section{Introduction}

New innovations in synchrotron data acquisition technology have led to an explosion of data being produced in synchrotron x-ray experiments~\cite{chupasSituXrayDiffraction2001,maierCombinatorialHighthroughputMaterials2007,chupasApplicationsAmorphousSiliconbased2007}.
%Where within the last decade, it was only possible to measure a single data point in a matter of hours.
%Now, such measurements can be performed in milliseconds~\cite{chupasSituXrayDiffraction2001,maierCombinatorialHighthroughputMaterials2007,chupasApplicationsAmorphousSiliconbased2007}
These developments allow researchers to study phenomena on much smaller time scales than ever before, opening up the possibility for the study of dynamic processes such as phase transitions and crystallization~\cite{kogermannXrayPowderDiffractometry2011}.
However, with this new speed of data generation, it is challenging to track changes in a material in real time.
For example, in an experiment where a phase transition is expected, the researcher may not have a good idea of when (or if) the phase transition occurred until they leave the synchrotron facility and finish analyzing the entire dataset at their home institution.

It would be much more informative if the experimental outcome could be monitored in quasi-real time and allow for adjustments in the measurement strategy or experiment design even during the beam-time. One challenge is simply to reduce raw data, for example, raw intensity images on 2D detectors, to something more scientifically meaningful on the time-scale of the experiment.
The second challenge is then monitoring the data to search for the scientific fact of interest.
The problem is tractable when most experiments at the beamline follow exactly the same pattern, in which case bespoke software can be engineered to give the user what they need.
This is the case, for example, at tomography beamlines~\cite{kakPrinciplesComputerizedTomographic2002,jacqu;nc13,jense;jes15} where every experiment is a tomography experiment, albeit on different samples.
However, many beamlines offer very flexible capabilities and the experimental workflow can differ considerably from one user to the next.
In this case we need flexible tools that can be reused but also easily adapted from experiment to experiment.
Also, in the context of an experiment, rather than a measurement, the outcome is not necessarily known and tools that allow the experimenter to explore their data in a flexible way are highly valuable.

An example is tracking of the phase evolution of a sample as it undergoes phase transitions, or during chemical synthesis.
In many cases the researcher does not have prior knowledge of the phases that will appear during the course of an \insitu\ experiment.
Therefore, it is critical to develop a model-independent method for monitoring the contributions and evolution of structural phases in a high-throughput x-ray diffraction experiment.
To that end, Principal Component Analysis (PCA), which is commonly used for data exploration~\cite{jamesIntroductionStatisticalLearning2013a}, was applied to study the growth of Mg(OH)$_2$ under different hydrothermal conditions~\cite{matosPrincipalComponentAnalysis2007} and to screen the crystallinity of Gd-based metal-organic framework (MOF) structures with a variety of organic backbones~\cite{lauPLUXterRapidDiscovery2011} and to recreate the constituent phases during the hydration process of Portland cement mortar~\cite{westphalUsingExploratoryFactor2015}.
%Matos et al showed that PCA can be used to classify samples of Mg(OH)2 prepared under different conditions into four distinct morphological groups based on growth conditions.
%Similarly, Westphal et al applied a PCA based method to a dataset containing the \textit{{\it in situ}} hydration of Portland cement mortar discovering that linear combinations of three principal components could be used to recreate the phases that were present within the process.
The PCA procedure was also shown to be effective in studying the evolution of phases present during an atomic pair distribution function (PDF) study of the nucleation and growth of Ag nanoparticles~\cite{chapmanApplicationsPrincipalComponent2015} and to access the amorphsivity induced into a crystalline drug using a cryo-milling technique~\cite{botkerAssessmentCrystallineDisorder2011}.
%Chapman et al used PCA to study dynamic processes. They discovered that while the raw components derived from the procedure were themselves not physical, taking linear combinations of them to recreate the PDF's baseline lead to representative PDFs of all the phases that were present in the experiment. Cole at al used PCA on PDF data to developed a regression algorithm  to predict the composition and material trends in rare earth phosphate glasses.

In the context of PCA, the principal components (PCs) are identified by finding linear combinations of the constituent data such that the variance accounted by each PC is maximized~\cite{jamesIntroductionStatisticalLearning2013a}.
In this process, the PCs are not expected to be related to signals that correspond to physical components of the system~\cite{longRapidIdentificationStructural2009,ermonPatternDecompositionComplex2015}, though they do accurately track how the physical components are varying with time as they encode variations in the signal.
Therefore, it is possible to determine the number of components in the set of data needed to account for most of the variations in the signal, and to see how they come and go as a function of time (if the dataset is a time-series), but it is difficult to get insight into the physical
origin of the changes, for example, which chemical species are appearing and disappearing during a chemical reaction.
%These holds to be especially true for PDF data, which is not as orthogonal as I(Q) data.
It is possible to carry out {\it post hoc} procedures to try and extract physical signals from the PCA outputs~\cite{westphalUsingExploratoryFactor2015,chapmanApplicationsPrincipalComponent2015,coleModelingPairDistribution2016a}, but these are generally somewhat bespoke, for example, by utilizing prior knowledge of the material being studied.

Non-negative matrix factorization (NMF) is an alternative data exploration technique~\cite{leeLearningPartsObjects1999}.
Similar to the PCA, the NMF seeks an approximation of the original dataset with a limited number of components. However, the input dataset and the extracted components are constrained to be strictly positive~\cite{paateroPositiveMatrixFactorization1994,leeAlgorithmsNonnegativeMatrix2001}.
The non-negativity constraint has been reported to result in more physical results in phase mapping from the x-ray microdiffraction data of an Fe–Ga–Pd ternary system~\cite{longRapidIdentificationStructural2009} and in determining phase diagrams and constituent phases from both x-ray diffraction and Raman spectra data on a thin-film with composition spreads in the Fe–Ga–Pd, (Bi,Sm)(Sc,Fe)O$_3$ and Fe–Nb–O systems~\cite{kusneHighthroughputDeterminationStructural2015}.
The NMF was also applied to a PDF study on accessing the degree of drug crystallinity and characterizing amorphous structural components in the PDF~\cite{s.geddesStructuralCharacterisationAmorphous2019}.  The challenge of NMF is that the matrix factorization procedure is not convex, and discovery of a global minimum solution is not guaranteed, but the preliminary studies cited suggest that it is a promising approach to the problem at hand.

Both PCA and NMF are defined for analysis of a fixed dataset which might be the result of a completed experimental campaign.
A challenge in the current context is the application of matrix factorization methods in the dynamic setting of an \insitu\ experiment.
In an automated experiment, or even more so in the context of a closed loop, autonomated experiment~\cite{taborAcceleratingDiscoveryMaterials2018b} where results of a previous experiment are guiding decisions about process parameters to use for subsequent measurements, we would like to follow the appearance and disappearance of reaction components or phase transitions in real time as the experiment progresses~\cite{billi;unpub20,rakit;jacs20}.
Therefore it is important to robustly extend the application of matrix factorization methods to this dynamic setting~\cite{todd;ic20}.
Here we present a computational infrastructure that addresses the two challenges of using matrix factorization methods in a streaming data context and obtaining more physical components, and tests it on sample data.

\section{Background}
\subsection{Introduction to PDF}

The atomic pair distribution function (PDF) $G(r)$ gives the scaled probability of finding two atoms in a material, a distance $r$ apart~\cite{egami;b;utbp12}.
Consider a structure with $N$ atoms and let $\{r_j\}_{j=1}^N$ denote the position of each atom.
The measured PDF $G(r)$ is defined as~\cite{farro;aca09,egami;b;utbp12}
\begin{equation}
\label{eq:sinft}
G(r)=\frac{2}{\pi}\int_{Q_{\min}}^{Q_{\max}} F(Q) \sin (Qr) dQ,
\end{equation}
where $F(Q)$ is defined as
\begin{equation}
\label{eq:fqmultielem}
F(Q) = \frac{1}{N\langle f\rangle^2}\sum_{l=1}^N \sum_{j\neq l}^{N}f_j^\star f_l\left(e^{-\frac{1}{2}\sigma_{jl}^2Q^2}\right)\frac{\sin(Qr_{jl})}{r_{jl}}.
\end{equation}
Here $F(Q)$ is the reduced structure function~\cite{egami;b;utbp12} and $Q$ is the magnitude of the scattering vector.
$r_{jl}$ is the distance between atoms $j$ and $l$.
$f_j$ is the form factor of the atom at position $r_j$, and $f^\star_j$ denotes its complex conjugate.
$\sigma_{jl}$ is the correlated broadening factor for the atom pair~\cite{proff;jac99,jeong;prb03}, which accounts for the thermal motions of atoms and other positional disorder.
%This mathematical model has been successfully used to study nanostructures by a number of authors~\cite{zhang2003water,cervellino2006efficient}.

\subsection{Principal Component Analysis and Non-negative Matrix Factorization}

Principal Component Analysis (PCA) and Non-negative Matrix Factorization (NMF) can both be described under a general framework of matrix factorization~\cite{paateroPositiveMatrixFactorization1994,leeLearningPartsObjects1999}.
A set of $m$ observed datasets of dimension $n$ can be expressed as an $n \times m$ matrix, $\vec{G}^{obs}$, where $n$ is the dimension of each observed data.
In this context, we are constructing an approximation of $\vec{G}^{obs}$
\begin{equation}
\label{eq:low_rank_approx}
\vec{G}^{obs} \approx \vec{G}^{e} \vec{W},
\end{equation}
where the dimensions of matrices $\vec{G}^{e}$ and $\vec{W}$ are $n \times p$ and $p \times m$ respectively and most importantly, $p < m$.
The $p$ columns of $\vec{G}^{e}$, denoted as $\vec{G}^{e}_p$, form the basis of this approximation and the $p$-th row of $\vec{W}$ represents the weight associated with $\vec{G}^{e}_p$.

The values for the matrices $\vec{G}^{e}$ and $\vec{W}$ that provide the best approximation in Eq.~\ref{eq:low_rank_approx} can be found by regarding Eq.~\ref{eq:low_rank_approx} as an optimization problem,
\begin{equation}
\label{eq:low_rank_approx_opti}
\min_{\vec{G}^{e}, \vec{W}} \norm{\vec{G}^{obs} - \vec{G}^{e} \vec{W}}_F,
\end{equation}
where $\norm{\cdot}_F$ indicates the Frobenius norm.
The difference between PCA and NMF resides in the constraints applied when solving Eq.~\ref{eq:low_rank_approx_opti}.

In PCA, the columns of $\vec{G}^{e}$ (also called principal components, PCs) are constrained to be orthonormal and the rows of $\vec{W}$ to be orthogonal to each other~\cite{leeLearningPartsObjects1999}.
Under this condition, the solution  can be obtained by carrying out a singular value decomposition (SVD) on $\vec{G}^{e}$,
which brings advantages in terms of computation speed.
Furthermore, the uniqueness of the solution for $\vec{G}^{e}$ and $\vec{W}$ is guaranteed~\cite{paateroPositiveMatrixFactorization1994,shlensTutorialPrincipalComponent2014}.

In NMF on the other hand, there is no constraint on the orthogonality but all the elements in $\vec{G}^{e}$ and $\vec{W}$ are constrained to be non-negative~\cite{paateroPositiveMatrixFactorization1994}.
This constraint makes finding the solution to Eq.~\ref{eq:low_rank_approx_opti} more challenging (the problem is not convex in general) and
despite advances in algorithms over time that make this feasible, the uniqueness of the solution is not guaranteed~\cite{leeAlgorithmsNonnegativeMatrix2001}.
On the other hand, it has been found that NMF generally gives  components, that could serve as a proxy to the actual measured signals in the data, whereas the PCA components are mathematically rigorous but bear little resemblance to physical signals~\cite{leeLearningPartsObjects1999,ermonPatternDecompositionComplex2015,stanevUnsupervisedPhaseMapping2018}.

\subsection{Streaming data analysis}

In the context of streaming data such as from \insitu\ experiments at a synchrotron beamline, the data arrive in a continuous, unbounded manner, as opposed to batch data where all the data is available at once.
This requires a different software architecture since each piece of data must be handled independently or carefully cached for future computation.
Here we make use of the \textsc{rapidz}~\cite{cw;20} Python library for streaming data analysis in
pipelines.  This package is an extension of the \textsc{streamz}~\cite{cw;streamz20} package.
It allows operation on the data streams, such as joining, branching and filtering, as well as allowing transformations to be applied to the events in the data stream computation.
Once a desired pipeline is set up, the data may be streamed through it.
The nodes in \textsc{rapidz} are highly customizable as they can be created based on general Python objects.

For PDF analysis, we have written streaming data analysis pipelines based on the xpdTools and xpdAn software packages, available in the xpdAcq Github organization (https://github.com/xpdAcq).
We used xpdTools to model the batch-NMF.
We then make use of xpdAn to turn the analysis capabilities of xpdTools into streaming mode.

Here we explore how to incorporate PCA and NMF into these pipelines, to provide a useful model-independent real-time monitoring of an \insitu\ experiment.
This presents special challenges because PCA/NMF are inherently batch processing algorithms since the decompose a static matrix of results.
We first demonstrate on a realistic simulated dataset the operation of PCA and NMF in batch mode.
We then discuss the results in the context of a real experiment.

\section{Simulated experiments}
\subsection{Simulated batch experiment}

\begin{figure}
	\includegraphics[width=0.7\columnwidth]{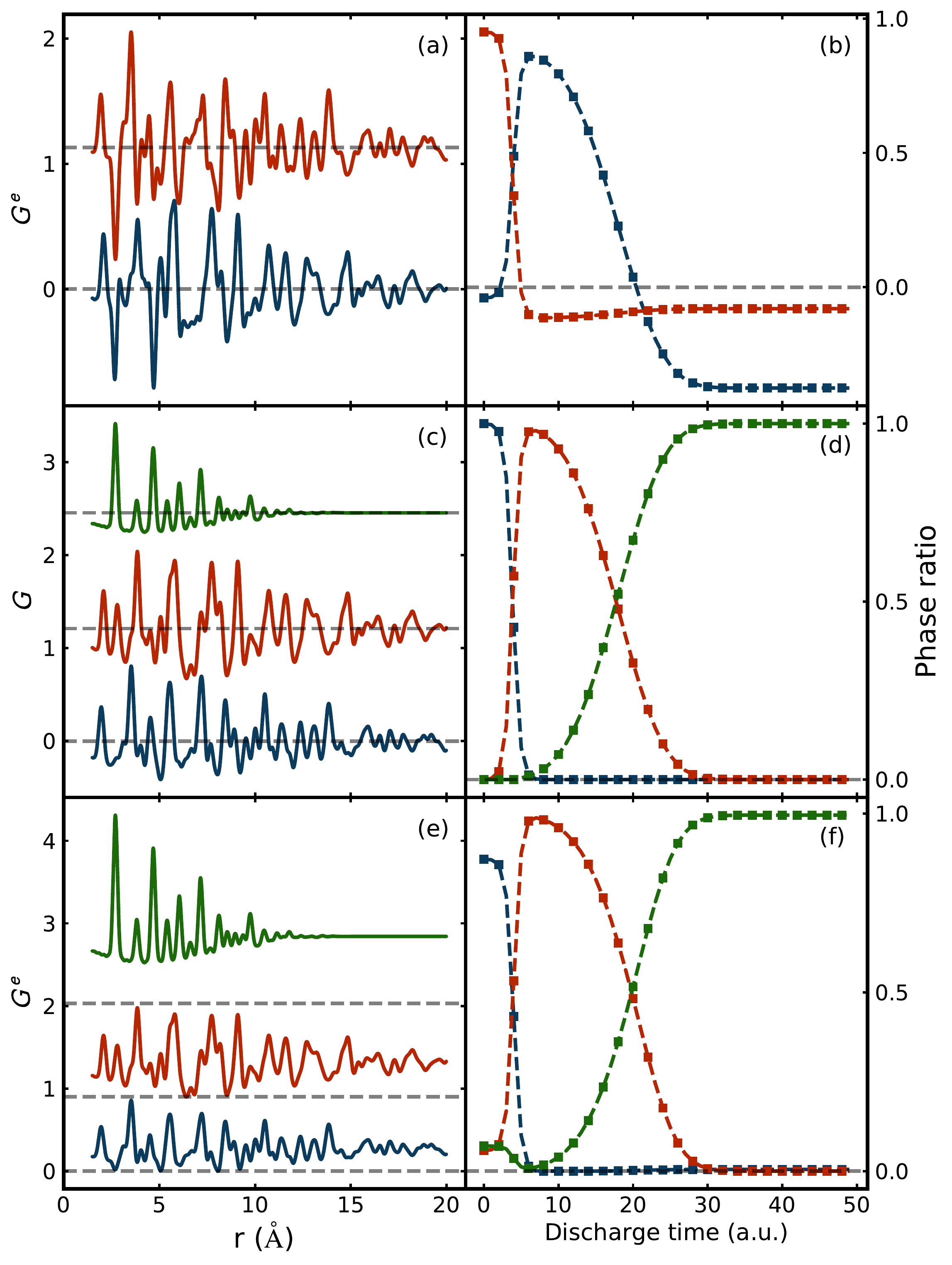}
	\label{fig:full_fig}
	\caption{Components of the signal and their weight coefficients as a function of discharge time in the simulated experiment.  The middle panel (c and d) shows ground-truth. The three PDFs (Ru nanoparticle (green), LiRuO$_2$ (red) and RuO$_2$ (blue)) that were linearly combined to make each PDF signal are in (c), and the time evolution of the weights of each component are in (d). the resulting signals are shown in Fig.~\ref{fig:sim_grs}. (a) and (b) show the result of carrying out PCA on the simulated data and (e) and (f) the result of carrying out NMF on the data.  PCA analysis (a),(b) only found two components when accounting for 99\% of the variance, shown in red and blue.  These components do not correspond to physical constituents.  On the other hand, NMF was quite successful at extracting signals that correspond to the physical constituents and their time dependence. The offset to each signal is plotted as the dashed line in the left panel.}
\end{figure}

\begin{figure}
    \includegraphics[width=0.7\columnwidth]{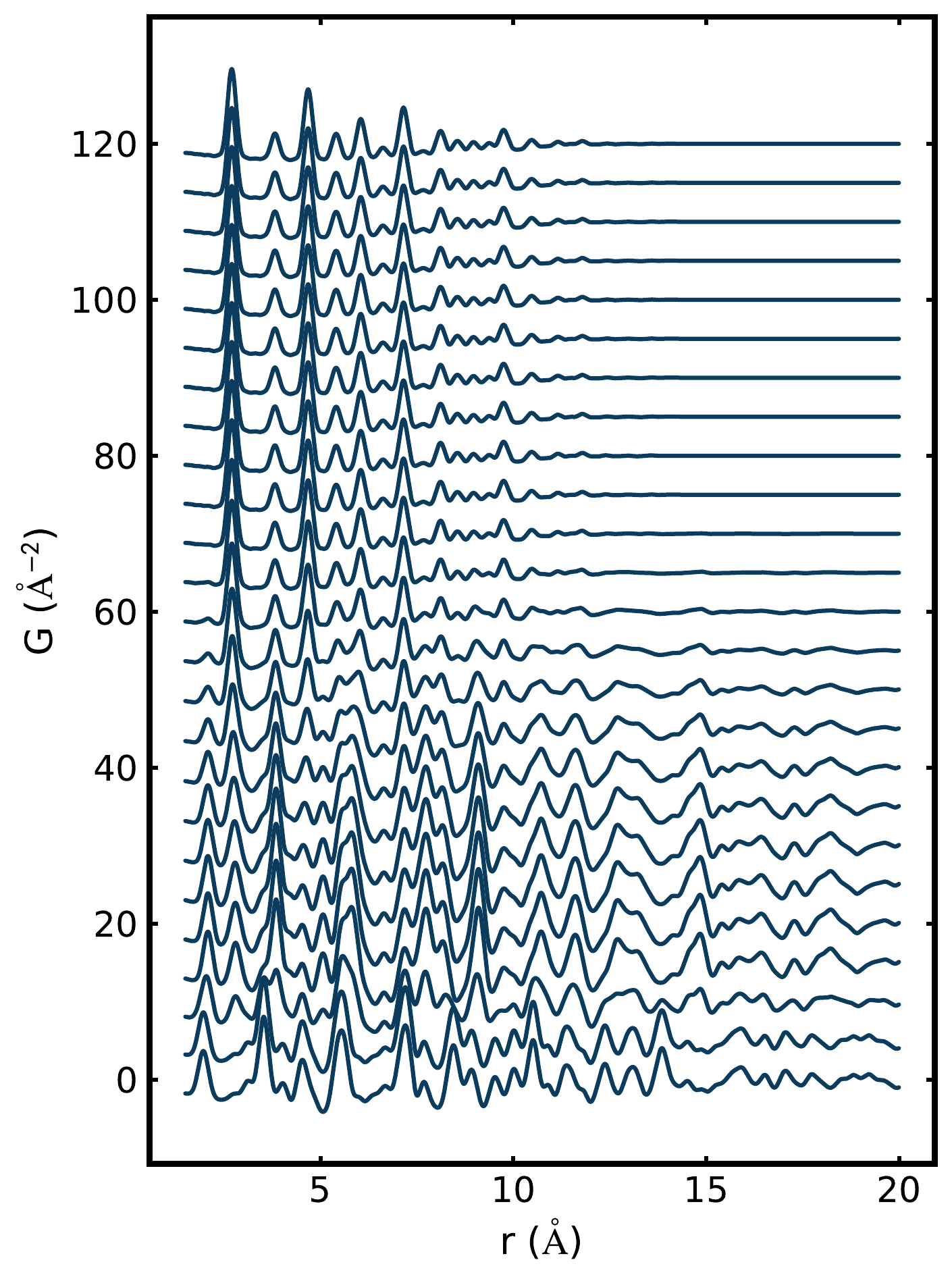}
    \label{fig:sim_grs}
    \caption{The simulated PDF signals obtained by making a linear combination of the PDFs of constituents (Fig.~\ref{fig:full_fig}(c)) with the weights shown in Fig.~\ref{fig:full_fig}(d). Note only every other PDF is shown here for ease of viewing.}
\end{figure}

To test the matrix factorization methods we here simulate 50 PDFs using phase mixtures of known PDFs so that we have a ``ground-truth" right-answer.
We then apply the matrix factorization methods to the simulated data.
To make the simulation somewhat realistic we consider the evolution of Ru-containing phases during the lithiation of a RuO $_2$ electrode that was first studied in~\cite{huOriginAdditionalCapacities2013},  where PCA was applied to the experimental PDFs in a batch-processed mode~\cite{chapmanApplicationsPrincipalComponent2015}.
The simulated PDFs each correspond to the progression of Ru-containing phases (RuO$_2$ $\longrightarrow$ LiRuO$_2$ $\longrightarrow$ Ru) identified during the electrochemical lithiation of RuO$_2$~\cite{huOriginAdditionalCapacities2013}.
The PDFs of the components were calculated using \cmi~\cite{juhas;aca15} from literature crystal structures for these phases~\cite{boman1970refinement,farleyInvestigationThermallyInduced1991,urashimaRutheniumNewMineral1974}, retrieved from the Inorganic Crystal Structure Database (ICSD)~\cite{belskyNewDevelopmentsInorganic2002}.
The simulated PDFs from these three constituent phases are shown in Fig.~\ref{fig:full_fig}(c).
The parameters for PDF calculation such as $Q$-range and instrumental resolution function are chosen to resemble realistic experiments and are reproduced in Table~\ref{tab:pdf_param} of Supporting Information section.
PDFs were then produced simulating the time dependence in the original experiment (Fig.~\ref{fig:sim_grs}) by making PDFs that are a linear combination of these phase components using the phase ratios shown in Fig.~\ref{fig:full_fig}(d).
Two matrix factorization algorithms, PCA and NMF, are carried out on the simulated dataset using the implementations in the \textsc{scikit-learn} python package~\cite{pedregosaScikitlearnMachineLearning2011a}.
In the our dataset of 100 runs, the PCA algorithm takes about 4.1~ms to factorize the simulated dataset while the NMF algorithm takes about 44.6~ms for the same task, both on the same machine with a single Intel i7-4700MQ CPU.

The PCA is closely related to the eigendecomposition of the variance-covariance matrix from the input data~\cite{jamesIntroductionStatisticalLearning2013a}.
The ratio of a eigenvalue to the sum of all eigenvalues yields the proportion of variance accounted for in the corresponding eigenvector ($\vec{G}^e$ in Eq.~\ref{eq:low_rank_approx}).
Empirically, we set a threshold of 99~\% of variance in the simulated data to determine the number of components to consider (the scree plot is reproduced in the SI in Fig.~\ref{fig:recon_error}).
With this setting, two components $\vec{G}^e$ are identified by the PCA algorithm (Fig.~\ref{fig:full_fig}(a)) and their weights at each simulated time-point are reported in Fig.~\ref{fig:full_fig}(b).
It is interesting that only two components are returned by the PCA given that the PDFs in the matrix were formed from 3 physical components.
We note that in PCA the components do not necessarily represent physical constituents but mathematical eigenvectors of the matrix, where the PCA requires the components to be orthogonal but allows the signals in the eigenvectors, and the weights, to take negative values, which is not possible for real physical components.
In general, PCA finds solutions that remove variance as rapidly as possible with the addition of each eigenvector, minimizing the number of eigenvectors needed to explain a certain level of variance, and it can't be used as a guide for how many physical components are present in the sample if those physical components have structural similarities.

On the other hand, the identified components $\vec{G}^e$ and their weights based on a non-negative matrix factorization are shown in Fig.~\ref{fig:full_fig}(e) and Fig.~\ref{fig:full_fig}(f).
The number of components for the NMF algorithm is determined by monitoring the falloff of the reconstruction loss reported in Fig.~S\ref{fig:recon_error}.
Somewhat surprisingly because it is a purely mathematical decomposition with no knowledge of chemistry, the NMF analysis identifies three components that closely resemble the simulated PDFs of the three Ru-containing phases (ground truth), with changes in relative scales between identified components similar to the ground truth.
The Pearson correlation coefficient between the identified components and three input phases (RuO$_2$, LiRuO$_2$, Ru) are 0.99, 0.99 and 0.97 respectively.
%The weights extracted by NMF yield a good agreement with the ground truth except for a small mismatch at the beginning of data series.

\subsection{Streaming PCA/NMF on simulated data}

Since one of the goals of this work is explore using matrix decomposition approaches to monitor \insitu\ PDF data at a synchrotron, we explore putting the PCA and NMF into a streaming data environment.
The first challenge here is to have the data be analyzed from raw images to PDFs on the fly.
At the XPD and PDF beamlines at the National Synchrotron Light Source II facility at Brookhaven National Laboratory, we have running a real-time streaming data analysis software framework using the \xpdacq and \xpdan packages.
These handle streaming data with the aid of the  Python based \rapidz streaming data framework.
We have therefore incorporated the PCA and NMF algorithms from \sklearn~\cite{pedregosaScikitlearnMachineLearning2011a} into a data analysis pipeline built using \rapidz.
To test the behavior of PCA and NMF in this context we stream the simulated PDF from our ground-truth dataset through the pipeline.

\begin{figure}
	\includegraphics[width=1\columnwidth]{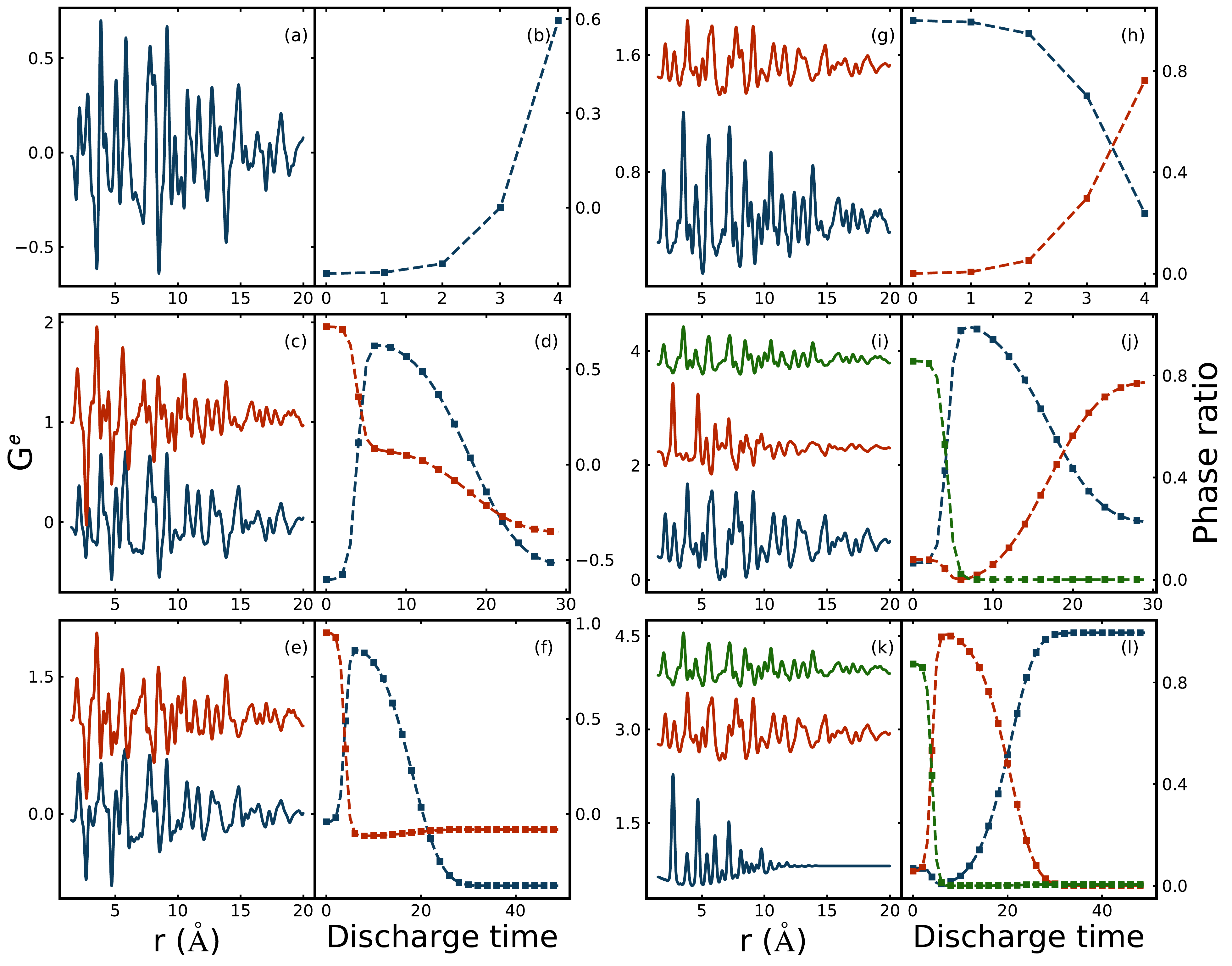}
	\label{fig:static_decom}
	\caption{Snapshots of PCA (a to f) and NMF (g to l) decompositions running in streaming mode. From top to bottom, each row shows the components identified by the algorithm (left panel) and the extracted weight ratio (right panel) when the algorithm is presented with the first 5, 30 and 50 simulated PDFs, respectively.}
\end{figure}

\begin{figure}
	\includegraphics[width=0.7\columnwidth]{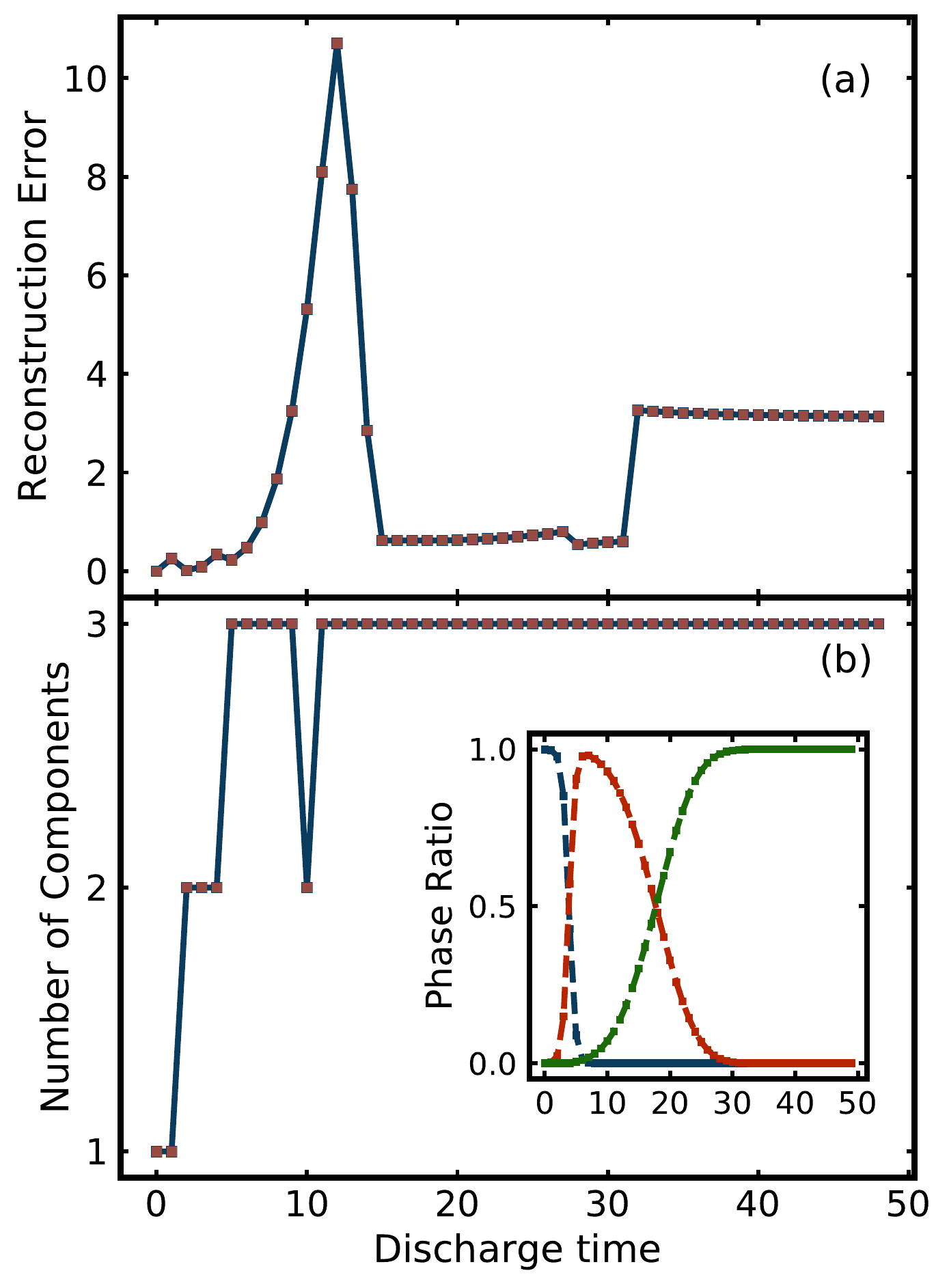}
	\label{fig:nmf_recon_error}
	\caption{(a) Reconstruction error and (b) the number of components for NMF with the simulated data as they arrive in streaming mode. The simulated phase ratio (ground-truth) is reproduced in the inset of (b)}
\end{figure}

The animations of this process are available both as part of Supplementary Information and online supporting material (\demourl).
Snapshots of the streaming NMF and PCA analysis are shown in Fig.~\ref{fig:static_decom}.
The evolution of the NMF reconstruction error and the number of components identified as the data arrive in streaming mode is plotted in Fig.~\ref{fig:nmf_recon_error}.
In this implementation, the PCA/NMF analyses are unaware of the totality of the components they will need at the end, so in the early stages of the experiment the detect fewer components.
When a new component is detected, the decomposition adds it, producing a discontinuous change.
This is useful and helpful information for the experimenter, but can be confusing.  at this point in time, in general, the existing components can reorder (and therefore abruptly change color and position in the visualization plots) which can be quite confusing.
Using a distance metric, such as the Pearson correlation coefficient, one might design an algorithm that makes a best-effort determination of which component is which and attempts to retain the order and the color of the existing components as a function of time, though results cannot be guaranteed.

Overall, the approach is successful and the use of PCA (which is faster) and NMF (which is slower but more physically meaningful in general) in a streaming environment looks to be a potentially very powerful capability.

\subsection{Batch and streaming PCA/NMF on experimental data}

Finally, we apply the methodology to actual experimental PDF data.
These are the same data that were reported in~\cite{huOriginAdditionalCapacities2013} and which we simulated above.
They correspond to the electrochemical lithiation of an RuO$_2$ electrode measured operando during charging.

\begin{figure}
	\includegraphics[width=0.7\columnwidth]{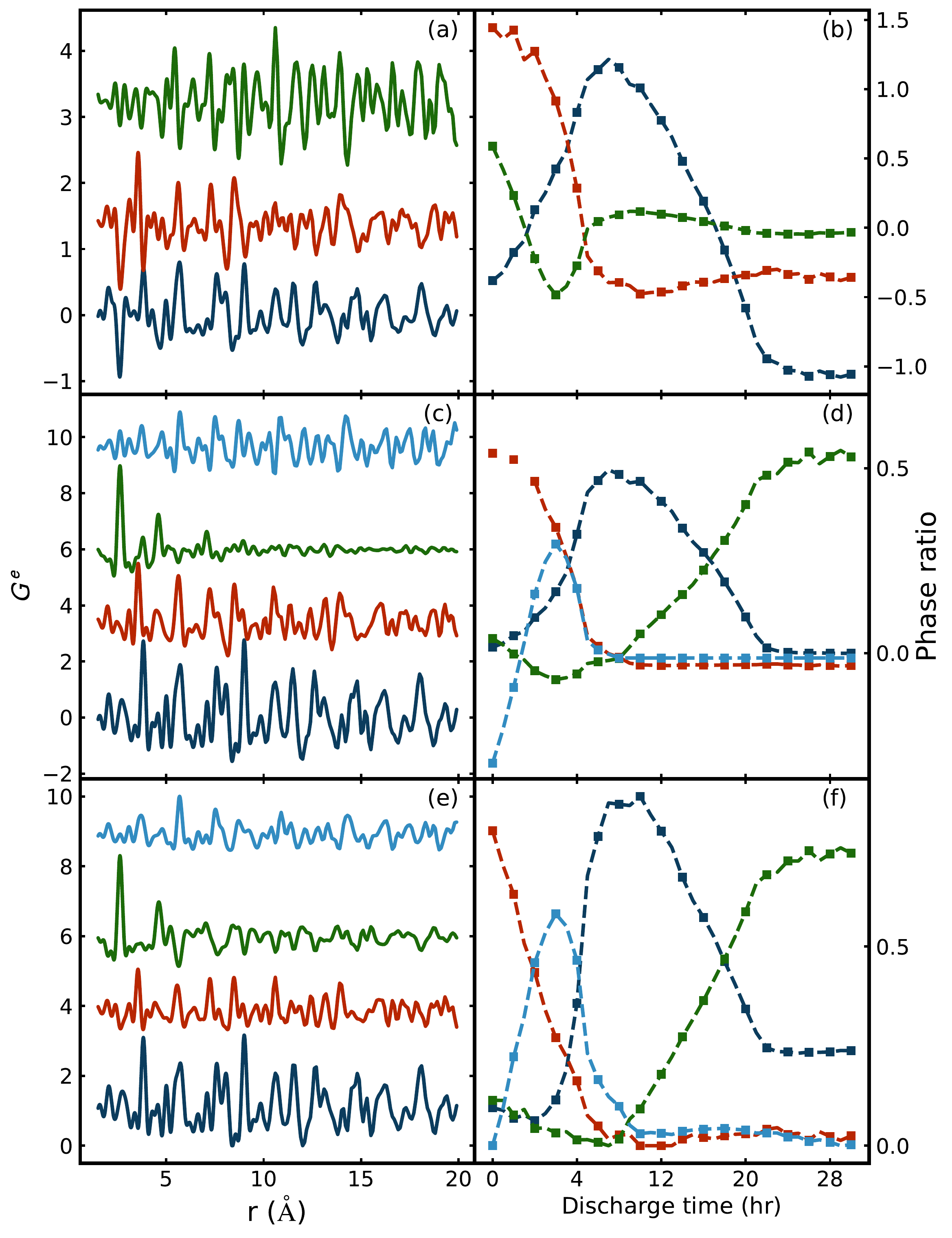}
	\label{fig:exp_full_fig}
	\caption{Components of the signal and their weight coefficients as a function of discharge time from the experimental data. The middle panel (c and d) show previously published results~\cite{chapmanApplicationsPrincipalComponent2015}. The components reported in (c) were obtained by taking linear combinations of PCA components following the rules summarized in the text. These components are referred to as chemical components (CCs) in the text. (a) and (b) show the result of carrying out a PCA analysis on the experimental data and (e) and (f) the result of carrying out an NMF on the experimental data.  PCA analysis (a),(b) only found three components when accounting for 99\% of the variance, shown in red, blue and green and these components do not correspond to physical constituents. On the other hand, NMF was quite successful at identifying signals that are close to the CCs and their time dependence.}
\end{figure}

\begin{figure}
	\includegraphics[width=0.7\columnwidth]{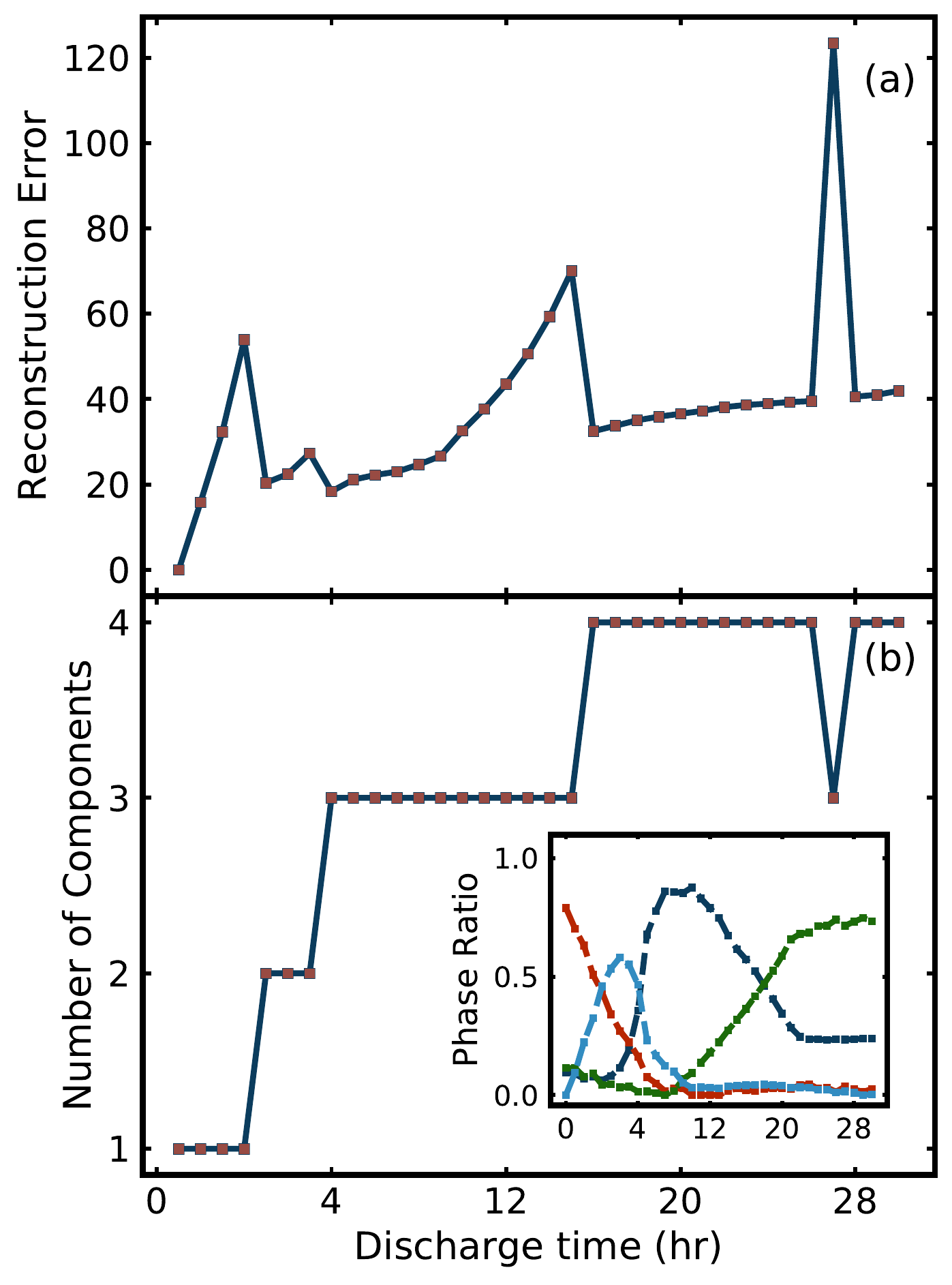}
	\label{fig:exp_nmf_recon_error}
	\caption{(a) Reconstruction error and (b) the number of components for NMF with the experimental data as they arrive in streaming mode. The identified phase ratio by NMF is reproduced in the inset of (b)}
\end{figure}

The \insitu\ synchrotron x-ray scattering experiments were carried out at the beamline 11-ID-B of the Advanced Photon Source, Argonne National Laboratory.
Details of the experiment and data reduction can be found in the original paper~\cite{huOriginAdditionalCapacities2013}.
The original analysis of the data first applied a PCA on the data followed by taking linear combinations of the PCA components in such a way as to remove negative peaks in the reconstructed components, abundances associated with the real phases were positive and the population of starting materials decreased monotonically while the population of final products increased population monotonically.~\cite{chapmanApplicationsPrincipalComponent2015}.
The resulting components are shown in Fig.~\ref{fig:exp_full_fig}(c), and the time evolution of the weights of these components are shown in Fig.~\ref{fig:exp_full_fig}(d).
The evolution of the NMF reconstruction error and the number of components identified as the experiment data arrive in streaming mode are plotted in Fig.~\ref{fig:exp_nmf_recon_error}.
For completeness we show a PCA analysis of the data in Fig.~\ref{fig:exp_full_fig}(a) and (b).
The NMF decomposition analysis is shown in Fig.~\ref{fig:exp_full_fig}(e) and (f), with the resulting NMF components in (e) and the time evolution of their weights in (f).
These reproduce the previous analysis rather well.
In the previous work, linear combinations of the PCA components were taken to agree with expected chemical components and we refer to these as ``chemical components".
To see how close the NMF components are to the previously reported chemical components we computed the Pearson correlation coefficient between them, over the range $1.5~\mathrm{\AA} \le r \le 20~\mathrm{\AA}$, as shown in Table~\ref{tab:pearson_exp_nmf_comp}.
\begin{table*}
    \centering
	\floatcaption{Pearson correlation coefficients between the NMF components and previously published ``chemical components" (CC). The chemical nature of CC identified in~\protect\cite{chapmanApplicationsPrincipalComponent2015} is reported in last column.All correlation coefficients $> 0.8$ are bolded.}
    \label{tab:pearson_exp_nmf_comp}
	\begin{tabular}{l|ccccc}
		& NMF 1               & NMF 2               & NMF 3               & NMF 4   & Chem. nature\\
		\hline
		CC 1 & {\bf 0.993}   & -0.132 & -0.323 & 0.496   & LiRuO$_2$ \\
		CC 2 & 0.0143 & {\bf 0.898}   & -0.0965 & 0.498 & RuO$_2$   \\
		CC 3 & 0.129  & -0.174 & {\bf 0.884}     & -0.0481 & Ru\\
		CC 4 & 0.226  & -0.244   & 0.0128 & {\bf 0.871}    & RuO$_2$ (distorted)
	\end{tabular}
\end{table*}
Notably, the components found by the non-negative matrix factorization procedure are each highly correlated with just one of the chemical components found in the original work, showing that the purely mathematical NMF with no human input found physically reasonable components from experimental data.

In an actual analysis of an \insitu\ experiment it might be of interest to discover the chemical nature of the components that are appearing and disappearing.
If they predominantly represent a single chemical phase it should be possible to identify the components by feeding them to a structure identifying algorithm such as the structure-mining approach~\cite{yang;aca20} that is
available as a web-service on the PDF in the cloud website~(\pdfitcurl)~\cite{ly;pdfitc20}.
We uploaded the extracted NMF components through this service.
Structure mining returned as its top-ranked structure LiRuO$_2$~(Pnnm), RuO$_2$~(P4\_2/mnm), Ru~(P6\_3/mmc) and LiRuO$_2$~(Pnnm) for components NMF1 through NMF4, respectively.
The structure found by the automatic searching algorith matches with the structures reported in the literature~\cite{huOriginAdditionalCapacities2013}.
It is interesting the structure-mining approach returns the same structure for NMF component 1 and component 4, where the latter is reported to be a distortion occurred during the transition from CC 2 (RuO$_2$) to CC 1 (LiRuO$_2$)~\cite{chapmanApplicationsPrincipalComponent2015}.

As noted from the tests of simulated ground truth data, the NMF components do not necessarily return actual PDFs of chemical components, but they are sufficiently close to them that it can result in a successful identification of the chemical species in the evolving reaction, on the fly without human intervention.

Finally, we simulated the NMF analysis of experimental data arriving in streaming mode.
The streaming approach worked in the same way as for the simulated data described in the earlier section.
The animations of streaming analysis are available both as part of Supplementary Information and online supporting material (\demourl).
The change of reconstruction loss versus the number of components for the experiment data is reproduced in Fig.~S\ref{fig:exp_recon_error}.

\section{Conclusion}
%We discuss the applications of two matrix factorization methods (PCA and NMF) for recognizing patterns of a diffraction data series under the context of traditional {\it post hoc} data analysis.
Here we validated the use of non negative matrix factorization for automatically extracting physically relevant components from PDF data. We
also studied the use of matrix factorization methods (PCA and NMF) in the context of streaming data coming from, for example, \insitu\ measurements.
We present a newly developed software infrastructure for doing this, and we demonstrate it on simulated and experimental PDF datasets obtained during the electrochemical lithiation of an RuO$_2$ electrode.
The streaming approach works well, capturing new components as they arise in the experiment.
By the end of the experiment the NMF decomposition recovers the result of traditional {\it post hoc} analysis.

The use of NMF for rapidly screening large numbers of time-series PDF data in real-time during an experiment looks very promising.
It may be use as a real-time diagnostic tool at the beamline in an \insitu\ experiment, but more importantly, may be used to carry out to on the fly decision making during and experimental campaign opening the door to active control of reactions, for example, or changing conditions once a reaction has clearly reached equilibrium.

\section{Acknowledgements}

This work was supported as part of GENESIS: A Next Generation Synthesis Center, an Energy Frontier Research Center funded by the US Department of Energy (DOE), Office of Science, Basic Energy Sciences under award No. DESC0019212.
Work done at Argonne and use of the Advanced Photon Source, an Office of Science User Facility operated for the US Department of Energy (DOE) Office of Science by Argonne National Laboratory, was supported by the US DOE under Contract No. DE-AC02-06CH11357.
\appendix
% This code ensures the supplementary captions have the right form
\setcounter{figure}{0}
\setcounter{equation}{0}
\setcounter{table}{0}
\makeatletter
\renewcommand\thefigure{\thesection\arabic{figure}} 
\renewcommand{\theequation}{S\arabic{equation}}
\renewcommand{\thetable}{S\arabic{table}}
\makeatletter
\section{Parameters for Simulated PDFs}
\begin{table*}
    \centering
	\floatcaption{Parameters used to calculate PDFs from atomic structures for the ground-truth dataset. All parameters follow the same definitions as in~\protect\cite{farro;jpcm07}.}
    \label{tab:pdf_param}
	{
		\begin{tabular}{c|c}
			Parameter & Value \\ \hline
			\rmin~(\AA) & 1.5\\
			\rmax~(\AA) & 20.0\\
			\qmax~(\AA$^{-1}$) & 25.0\\
			\qmin~(\AA$^{-1}$) & 0.1\\
			$Q_{damp}$~(\AA$^{-1}$) & 0.1 \\
			$Q_{broad}$~(\AA$^{-1}$) & 0.04 \\
		\end{tabular}
	}
\end{table*}
\section{Reconstruction loss of PCA and NMF}
\begin{figure}
    \includegraphics[width=0.7\columnwidth]{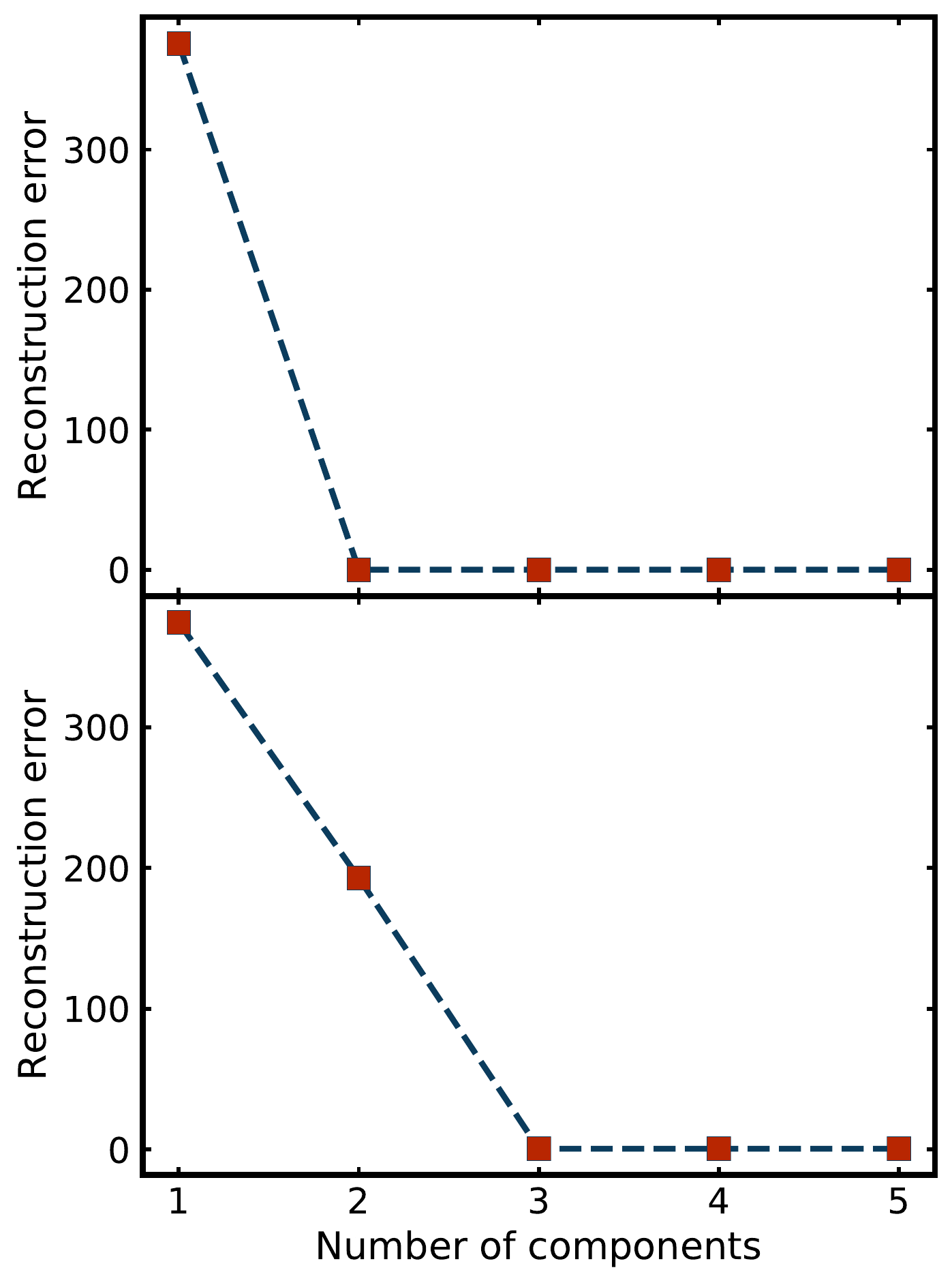}
    \label{fig:recon_error}
    \caption{The reconstruction error (defined in Eq.~\ref{eq:low_rank_approx_opti}) v.s. the number of components used for PCA (top) and NMF (bottom) algorithms on the simulated data.}
\end{figure}
\begin{figure}
	\includegraphics[width=0.7\columnwidth]{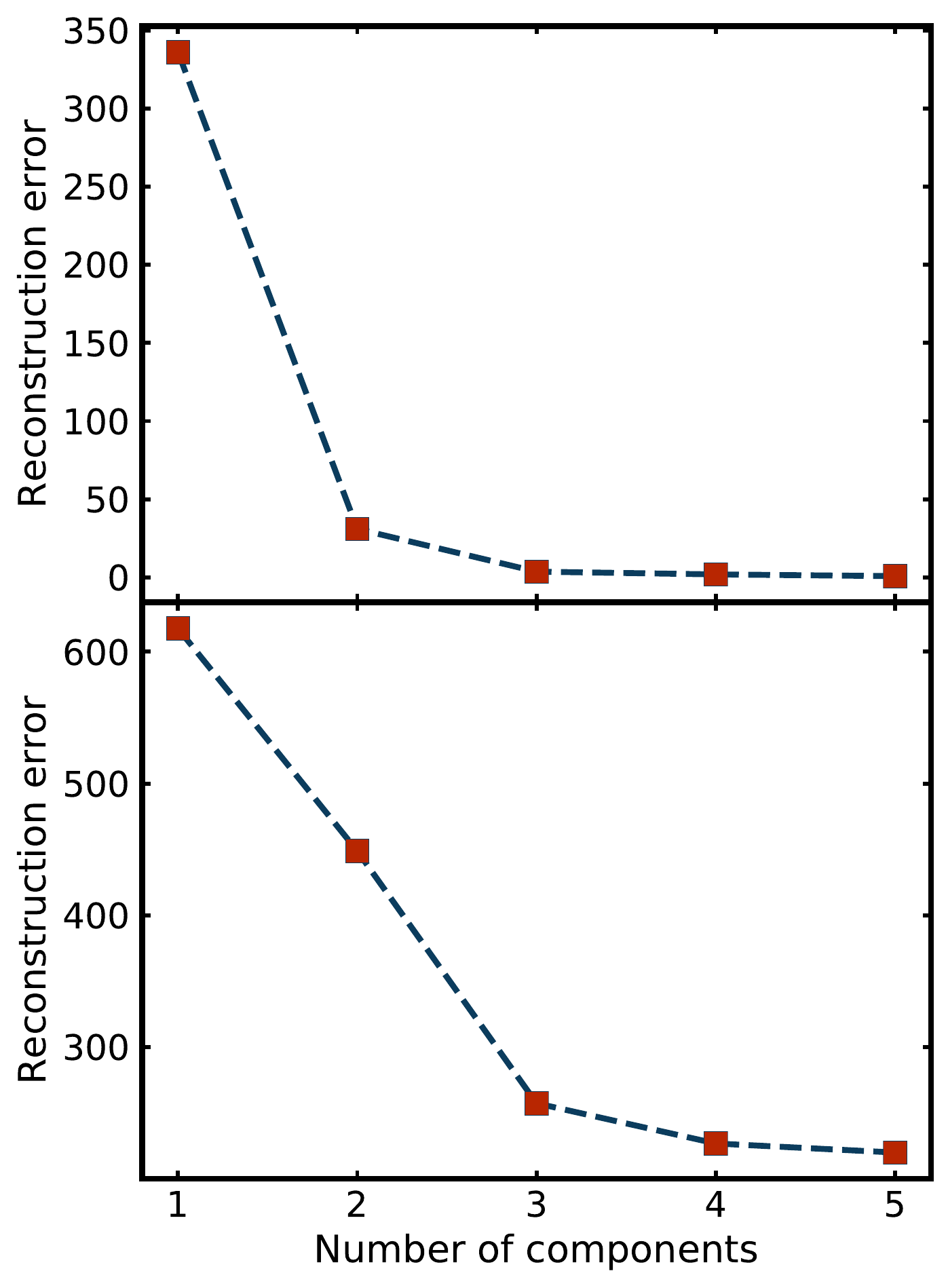}
	\label{fig:exp_recon_error}
	\caption{The reconstruction error (defined in Eq.~\ref{eq:low_rank_approx_opti}) v.s. the number of components used for PCA (top) and NMF (bottom) algorithms on the experiment data.}
\end{figure}
\bibliography{sbillinge,19clsab_NMF_PCA,19clsab_NMF_PCA_misc,ref}
\bibliographystyle{iucr}
\end{document}